# Security Analysis of Mobile Banking Application in Qatar

Shaymaa Abdulla Al-Delayel

Cyber and Network Security Department,

Information Technology Division,

Community College of Qatar

shayma.al-delayel@hotmail.com

**ABSTRACT**

This paper discusses the security posture of Android m-banking applications in Qatar. Since technology has developed over the years and more security methods are provided, banking is now heavily reliant on mobile applications for prompt service delivery to clients, thus enabling a seamless and remote transaction. However, such mobile banking applications have access to sensitive data for each bank customer which presents a potential attack vector for clients, and the banks. The banks, therefore, have the responsibility to protect the information of the client by providing a high-security layer to their mobile application. This research discusses m-banking applications for Android OS, its security, vulnerability, threats, and solutions. Two m-banking applications were analyzed and benchmarked against standardized best practices, using the combination of two mobile testing frameworks. The security weaknesses observed during the experimental evaluation suggest the need for a more robust security evaluation of a mobile banking application in the state of Qatar. Such an approach would further ensure the confidence of the end-users. Consequently, understanding the security posture would provide a veritable measure towards m-banking security and user awareness.

## 1 INTRODUCTION

Mobile devices, data security, and privacy has progressively become a significant worry in the field of technology with the development of devices and applications, which increases the risk of finding penetrations regarding the information and data provided by users [1]. Security and privacy are two cardinal challenges often associated with any ubiquitous device, especially in relation to the user of the system [2]–[4]. Assuring the end-user of the security and integrity of a mobile platform is essential a core functionality of any ubiquitous device. Mobile devices mostly rely on mobile applications. so, in the field of mobile banking, applications will benefit the user and the bank by letting the user communicate easily with the bank, affirmed that the extraordinary increment of mobile applications and its immediate effect on the security of client's device and the impact on the information should be taken seriously [5].

Banking has never been as secure and easy as today; thanks to technology. With the increase of mobile applications, banks have taken the opportunity to offer their services through a diverse application, where almost every bank has gravitated towards digitalization, to provide enhanced customer service, reduced errors, and managing a large volume of banking transactions. "Mobile banking means that users adopt mobile terminals to access various payment services" [6], such as money transactions, bill payment, financial management, account statement, available balance, and more services.

The world's first fully mobile banking application was launched in 1999 by the Bank of Scotland, where the bank offers its clients an opportunity to do some of the banking transactions over the internet using SMS service. As mobile banking applications developed over the years, the use has been easier and safer than before. Furthermore, it provides the users with a high-security method that will help the bank to get customer satisfaction. Also, some banks offer a feature that allows the clients to apply a money transaction through an m-banking application to an ATM through a QR code to withdrawing the money. This is one of the basic features of a mobile banking application [7].

In Qatar for instance, there are almost 26 banks but not all of them have an m-Banking application, whilst about 21 banks who have it, sandwiches between local and foreign owners. Each bank has almost the same services but only a few of them have an advanced service, such as, card-less E-cash money transfer service, which is a simple and secure way to send cash through an m-banking application. These features in the m-banking application, and the steady rate of adoption, it has effectively demonstrated how banks in Qatar have made life easier. Customers can enjoy the bank service while sitting at home, instead of wasting time waiting in the customer service line. These m-banking services are, however, faced with diverse security challenges. As asserted in a study in [1], mobile platforms present an easier threat target that can be used to conduct cyber-related attacks. This is particularly true when mobile devices are connected to cloud services which are focal points for cyber threat actors [8].

### 1.1 Problem Background

Mobile application security (the security layer) is one of the most important features for the developer and users of m-applications, without which the confidence of the users can be eroded. Mobile application security focuses on the software security of the mobile application in different platforms such as Android-OS, iOS, Harmony-OS as well as Symbian-OS. Nowadays, more users rely on a mobile application for many of their digital tasks over desktop applications. Although, these applications have access to a large amount of user data, which is considered sensitive information that the user wants to keep safe from unauthorized access, such as numbers, pictures, emails, documents, and more. Mobile applications are providing security control options that help developers in building a secure application. However, failing in providing the high-security features will lead to having vulnerabilities which will make the application get attack easily and affect the user's data. Some of the common issues in mobile applications might be a Mobile terminal domain, Access and transmission domain, failing to fulfill security requirements, and more [9].

Despite all the security measures in an m-banking application, there might be a lot of vulnerabilities, which affect the bank and users of the application [10]. "According to the recent MQA survey, security is a major problem in the adoption of m-banking and approximately 72% of respondents state that they are concerned about the safety of access to financial data on a mobile device" [11]. This study shows that many users of the m-banking application are not feeling safe to use it and it might put them at a risk that all their financial information saved on the mobile. Like any other applications, m-banking applications have various vulnerabilities that might refer to using a Wi-Fi connection and different operating systems which affect the security methodology on each mobile OS. Using mobile banking applications can be easy and could potentially save time. However, currently, there is no public information on the security of mobile banking applications in Qatar, which can be used to evaluate the security posture of any banking application in Qatar. This further implies that the metrics used to evaluate the security awareness by the end user remains unquantified.

## 2 LITERATURE REVIEW

In [5], research was done that focuses particularly on banking mobile applications of banks operating in Ghana, they follow a method to Analyzed how their respective official mobile applications were developed either as native android apps or cross-platform apps. The study further delved into the android security-based permission models used by selected banking mobile applications, and they used a cross-platform app identifier, a command-line program written in Python to decode an APK file. Also, they used AVC UnDroid, which is an online Android permission analysis experimental tool. The result that they found was All banking applications that were identified as using a cross-platform development framework used Apache Cordova. Out of the total of twenty applications, 9 representing approximately 40% of the applications used cross-platform development framework while the remaining 13 representing approximately 60% did not use any of the popular cross-platform development frameworks hence can be concluded as using native development framework.

Another research was done [12], which attempted to complement the former reviews by expanding the coverage of Android security issues, and malware growth, and follow a method that use Static and dynamic analysis to detect security-sensitive and malicious behaviors of apps, to do so the evaluation technique was to Implement a prototype to evaluate using many apps collected from different official and unofficial app markets, at the end a Comparison for results of the proposed solutions for addressing the permission management on Android OS, and found that more than a hundred replicas running on a single remote server have been made.

In [7], this research has been made to address the security risks in mobile and online banking, especially in mobile banking is a major problem for the banks and the users because of the innovations brought by the technology and security gaps in every innovation, by using NowSecure Lab Automated tool to assess the security threats in mobile banking to Examine security threats and security measures in mobile and online banking systems. As a result, they found that the security vulnerabilities in banking applications are searched by malicious people with various methods and they will continue to search for these security vulnerabilities. The process employed to evaluate the security of the mobile applications is presented in the next section.

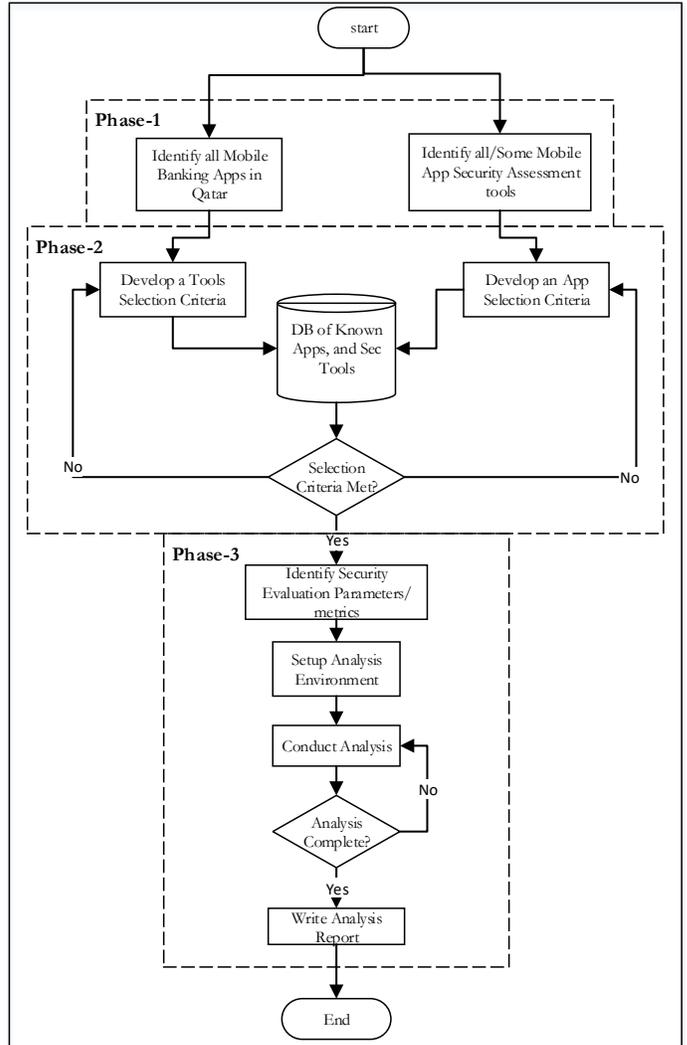

**Figure 1:** Research Operational Framework

## 3 RESEARCH METHODOLOGY

An overview of the research framework followed in this study is presented in Figure 1. The research framework comprises three inter-related phases which detail the analysis environment, the installation processes, as well as the appropriate download, required for each mobile application such as the APK. Such files contain useful information that can aid in conducting the analysis process. Then a selection criterion for the applications and tools used in this study is provided. Lastly, setup steps to develop the analysis environment and analysis result comparison are provided. Further details of these processes are presented in the subsequent subsections.

## 3.1 Phase 1: Preparatory Stage

Initially, there are some tools to do the security analysis of selected android applications. Some of the tools are paid, some are online, some of them only provide support at the development level and a small number of them are available for open-source security analysis as well as support post-deployment mobile applications packages. After exploring those tools that are open-source and can perform static analysis on the required mobile applications.

Besides, Ubuntu OS is installed, as most of the professional and security researchers prefer to use this as compared to Windows for security analysis as it the most secured OS. The primary reason for using Linux is that most of the available security tools are Linux-based. Moreover, the support on technical forums for such tools is available on Ubuntu. To install Ubuntu as a second OS on the used machine which is one of the commonly used methods to run two OS on a single machine as dual boot (Windows along with Ubuntu, Parrot security, or Kali Linux) and install a GRUB (GR and Unified Bootloader) on the machine for OS selection.

Then APK's files of the required banking applications should be extracted from the Internet. Apps can be download and install from the Play Store and extract their APK's from internal storage located at the following path /data/app/, but it is only accessible if the android mobile is rooted. If the mobile is not rooted APK files can install using third-party apps from Play Store (such as APK Extractor) that gets the desired APKs installed on the mobile device. After getting the APK's on mobile, a copy of them should be on a laptop or personal computer for further analysis. Moreover, there are multiple websites available on the internet that provides the APK's of the android apps available on the Play Store such as apk4fun.com, apkpure.com, and apkmirror.com, etc.

## 3.2 Phase 2: Selection Criteria

Each application and tool are chosen based on specific selection criteria that will fit the scope of the study.

*Selection Criteria for apps*

The primary motive behind taking these two apps as a case study is that both banks are famous and widely used by the people of Qatar and both banks make their way up to the top ten banks of Qatar. QIB and QIIB android apps are chosen because of the popularity, service, and security provided to the end-users. Furthermore, QIB has some progressive methods over QIIB so which would make the comparison and understandability of the report easier for the readers of this document. The QIB and QIIB mobile applications are hereinafter referred to as m-banking app-1 and m-banking app-2 respectively.

*Selection Criteria for tools*

Consequent to the above discussion, open-source tools are chosen, those tools work seamlessly with the APK packages. Following are the tools that were shortlisted for the activity. A summary of the comparative analysis of known open-source tools for Mobile application security evaluation is further presented in Table 1.

- QARK (Quick Android Review Kit)
- MobSF (Mobile Security Framework)

Reasons for choosing these tools are they fit the above criteria. MobSF is an all-in-one mobile apps pen-testing framework proficient in executing static, dynamic, and malware analysis. It is hosted in a local environment, so sensitive data never interacts with the cloud. While QARK is a tool designed to identify potential security vulnerabilities of java-based Android mobile apps. It is designed explicitly to demonstrate the potential issues in them.

**Table 1**: Comparative analysis of Mobile application security evaluation tools

| Tools | Platform | Sources | Benefits | Limitations |
|---|---|---|---|---|
| Appium | Android iOS Hybrid | Open Source | • It supports variety of languages (JSON wire protocol).<br>• It supports both Android & iOS.<br>• It supports the automation of hybrid and web-apps also.<br>• It does not need access to source code.<br>• It supports different frameworks.<br>• APIs can be integrated.<br>• It has all the functionality of selenium.<br>• It has support for built-in applications like camera, calendar, and phone. | • It does not support image comparison.<br>• Long time to configure for both android and iOS.<br>• Appium documentation is a little weak.<br>• A smaller amount of available tutorials.<br>• It has a deficiency of record and plays functionality. |
| Robotium | Android | Open Source | • It is a library for unit tests.<br>• It supports a variety of versions and subversions of Android.<br>• The run-time binding to GUI components makes the test cases more robust.<br>• It enables users to test more flexibly for scrutinizing results.<br>• Very good readability of test cases.<br>• Fast execution of test case.<br>• Automatic timing and delays. | • It only supports Android.<br>• It only supports Java.<br>• It doesn't support web-apps.<br>• It has a deficiency of record and plays functionality. |
| Selendroid | Android iOS Hybrid | Open Source | • It supports no of languages i.e. C#, Java, Python, Perl, etc.<br>• No need to change the application to automate the test.<br>• It supports the hot plugging-in of physical devices.<br>• A test can be run on multiple physical devices and emulators.<br>• Has an inspector to make simple the test-case development. | • Supports only limited gestures.<br>• Supports for only Android < 4.1<br>• It doesn't support the Toast Messages.<br>• Built-in apps i.e. maps, cameras, etc. can't be automated. |
| Monkey Runner | Android | Open Source | •Automate functional tests for Android apps.<br>•No need to know the source code while writing test cases.<br>•It can run a test on real devices and emulators concurrently.<br>•It has a recording tool for creating test cases.<br>•Enable user to control the devices with code outside of Android code | • Every device has a different test case script.<br>• User has to change the script after every change accordingly.<br>• Insufficient documentation. |
| Calabash | Android iOS Hybrid | Open Source | •It has a huge and passionate community for support.<br>•It has support for Emulators, so no need for physical devices.<br>•Test cases are easy to write because it supports natural language.<br>•It supports all the simple events and movements present in the libraries.<br>•It can query the Web-Views with CSS selectors | • When a step in the test is failed then the successive are skipped.<br>• No IDE or Editor is available.<br>• It is very friendly towards the Ruby.<br>• Apple doesn't support this project, so functionalities alter ominously from one OS to another |
| KIF | iOS | Open Source | •It minimizes the in-direction means all of the test case are developed in Objective-C.<br>•It integrates directly with X-Code project.<br>•It has wide iOS coverage i.e. from iOS which implies you've to recover 5.1 to iOS 9.<br>•It imitates the real user input like tap events and delete gestures.<br>•It is built-in in X-Code 5. User can run KIF test using Test Navigator.<br>•It can keep the track of all activities. | • Just for iOS, no Android version.<br>• Some functionalities may be buggy.<br>• It can't recover failure, which means you've to recover the app from failures. If you don't then there will be a domino-effect of failures. |
| Monkey Talk | Android iOS Hybrid | Open Source | •It is easy to use and setup.<br>•Same test script for different platforms.<br>•It has good support for gestures.<br>•No need to change the app to automate the test.<br>•No programming proficiency is required.<br>•It can generate HTML and XML reports. | • It supports only command language or JavaScript API.<br>• It does not support the HTML5 web-apps.<br>• It requires access to the source code of the application. |

## 3.3 Phase 3: Setup and Analysis

Assuming Ubuntu is installed. The following setup is mandatory to install the required tools for analysis. Since we have to use two tools, so install both of them sequentially. Starting with MobSF (Mobile Security Framework) setup.

i. Download and install prerequisites packages via the command line.
ii. Download and install MobSF via command line from git.
iii. Open http://localhost:8000 using the browser.
iv. Uploading the required apk to MobSF.
v. Scan; which will take some time.
vi. A report; will be generated by MobSF to pick out the finding.
vii. Conclude and draw results from the reports and add them to the report.

Setup up for QARK (Quick Android Review Kit)
i. Install an Ubuntu or OSX.
ii. Download and install prerequisites packages via command line.
iii. Clone the repository of QARK.
iv. Verify QARK installation.
v. Navigate to the QARK directory and run it via terminal.
vi. Type the command that will instruct QARK to initiate the de-compilation process and to begin the static code analysis on the decompiled code.
vii. QARK generates an HTML report for the static analysis that will be helpful to conclude.

## 4 RESULT

The result of the exploration process using the methodology discussed in the previous section is presented in this section. More specifically, on how to perform a security evaluation analysis of selected mobile banking apps. The summary result of the security analysis tools: MobSF and QARK, is presented in Table 2. Both security tools are open-sourced and popular tools for the static analysis widely used to identify the potential security vulnerabilities of an android application. Using the CVSS metrics, both mobile applications have an average CVSS of 6.2 and 6.6 respectively. A common vulnerability scoring system (CVSS) is an industry standard metrics used to rank the observable severity of a vulnerability of a system under investigation. CVSS ranges from extremely severe (10) to least severe (0). A typical secure mobile application is expected to rank between 1 and 3. This result does reflect a poor security posture for the explored m-banking application. Furthermore, both applications ranked 10/100 in terms of security score. A value of 10, as observed in this study is considered critically poor, which further substantiate the CVSS rating. The m-banking app-1 integrates external storage, hardcoded URLs, insecure functions, a broadcast is sent without receiver permission, logs in the complied in the app, vulnerability ECB cipher usage, empty certificate method, backup is allowed in manifest and potentially vulnerable check permission function is called. Similarly, the m-banking app-2 exhibit similar poor security posture. A further evaluation of the distinctive CVSS for each component further reveal that there are fundamental hotfixes which have been reported in the security community in the common weakness enumeration (CWE) and OWAPS top 10. The lack of standardized security practice is alarming. For instance, the storage of sensitive information in a cleartext format was observed in both m-banking application. The

MD5 hashing algorithm is considered to have a poorer security posture, which posit that it should not be implemented in critical or sensitive information. Furthermore, the use of an insecure random number generating to enhance the security of the m-banking app is a poor choice. Given that m-banking application could be a potential threat vector for the banking system, the deployment of such weak security apparatus further depicts the security posture.

Table 2: Observed Security Posture of the Mobile Banking Applications

| No. | Issue | Severity | Standards |
|---|---|---|---|
| 2 | The App uses an insecure Random Number Generator. | High | CVSS V2: 7.5 (high). CWE: CWE-330 Use of Insufficiently Random Values. OWASP Top 10: M5: Insufficient Cryptography. OWASP MASVS: MSTG-CRYPTO-6. |
| 4 | Files may contain hardcoded sensitive information like username, password, keys, etc. | High | CVSS V2: 7.4 (high). CWE: CWE-312 Cleartext Storage of Sensitive Information. OWASP Top 10: M9: Reverse Engineering. OWASP MASVS: MSTG-STORAGE-14 |
| 5 | SHA-1 is a weak hash known to have hash collisions. | High | CVSS V2: 7.4 (high). CWE: CWE-312 Cleartext Storage of Sensitive Information. OWASP Top 10: M9: Reverse Engineering. OWASP MASVS: MSTG-STORAGE-14 |
| 7 | The app can read\write to External Storage. Any App can read data written to External Storage. | High | CVSS V2: 5.5 (medium). CWE: CWE-276 Incorrect Default Permission. OWASP Top 10: M2: Insecure Data Storage. OWASP MASVS: MSTG-STORAGE-2 |
| 8 | MD5 is a weak hash known to have hash collisions. | High | CVSS V2: 7.4 (high). CWE: CWE-327 Use of a Broken or Risky. OWASP Top 10: M5: Insufficient Cryptography. OWASP MASVS: MSTG-CRYPTO-4 |
| 9 | The app creates a temp file. Sensitive information should never be written into a temp file. | High | CVSS V2: 5.5 (medium). CWE: CWE-276 Incorrect Default Permissions. OWASP Top 10: M2: Insecure Data Storage. OWASP MASVS: MSTG-STORAGE-2 |

Table 3: Application status for M-Banking App-1

| Dangerous App Permissions of m-banking app-1 | Status |
|---|---|
| Com.google.android.finsky.permission.BIND_GET_INSTALL_REFERRER_SERVICE | Dangerous |
| Android.permission.WRITE_EXTERNAL_STORAGE | Dangerous |
| Android.permission.READ_PHONE_STATE | Dangerous |
| Android.permission.READ_EXTERNAL_STORAGE | Dangerous |
| Android.permission.READ_CONTACTS | Dangerous |
| Android.permission.CALL_PHONE | Dangerous |
| Android.permission.ACCESS_FINE_LOCATION | Dangerous |
| Android.permission.ACCESS_COARSE_LOCATION | Dangerous |

Table 4: Application status for M-Banking App-2

| Dangerous App Permissions of M-Banking App-2 | Status | Information |
|---|---|---|
| Android.permission.ACCESS_COARSE_LOCATION | Dangerous | Coarse (network-based) location. |
| Android.permission.ACCESS_FINE_LOCATION | Dangerous | Fine (GPS) location. |
| Android.permission.CAMERA | Dangerous | Take pictures and videos. |
| Android.permission.GET_ACCOUNTS | Dangerous | List accounts. |
| Android.permission.READ_CALENDAR | Dangerous | Read calendar events. |
| Android.permission.READ_EXTERNAL_STORAGE | Dangerous | Read external storage contents. |

On the other hand, m-banking app-2 has the following issues, vulnerability ECB cipher usage, JavaScript enabled in web-view, vulnerable check permission function, logs in the complied in the app, insecure functions, a broadcast is sent without receiver permission, external storage, web-view enables file access, remote debugging enabled in web-view and empty pending intent found and many more are briefly defined in this report in detail. All these vulnerabilities raise many questions on user privacy and can lead to major security breaches at any time. Both Banks should pay more focus on the security of their applications and need to fix the following issues explained in this report as soon as possible, so that both consumers and stakeholders do not have to face any consequences shortly.

## 5. DISCUSSION

As the result indicates that both banking apps did not pass the baseline security requirement and have a lot of vulnerabilities and weaknesses, which makes the security measures of both mobile banking applications are at a low level. Starting with the list of m-banking app-1 vulnerabilities such as usage of mobile external storage, hardcoded URLs, insecure functions, a broadcast is sent without receiver permission, logs in the complied in the app, vulnerability ECB cipher usage, empty certificate method, backup is allowed in manifest and potentially vulnerable check permission function is called. The m-banking app-2 is no more behind in term of weak security and have much more serious vulnerabilities as compared to m-banking app-1 such as vulnerability ECB cipher usage, JavaScript enabled in web-view, vulnerable check permission function, logs in the complied in the app, insecure functions, a broadcast is sent without receiver permission, external storage, web-view enables file access, remote debugging enabled in web-view and empty pending intent found. Both m-banking applications have a lot of vulnerabilities that could lead to a serious security breach to their customers and invade their privacy. The results demonstrate that in Qatar, android based banking applications have a lot of security vulnerabilities and weaknesses that hackers can exploit to gain access and cause serious damage to the reputation of banking organizations. The research and security community can get a brief and in-depth summary of the weakness and vulnerabilities of the m-banking app-1 and m-banking app-2 android-based applications and offer help organizations to fix these issues at the earliest Furthermore, the research community can get a brief overview of the security warnings, loopholes, and vulnerabilities of m-banking application within the State of Qatar from this study. Consequently, this knowledge can provide a baseline for further exploration of the security weaknesses in Android as well as iOS banking applications in Qatar. It would make a great impact on the society and put the great trust of people in banking organizations of Qatar. As a result, people could feel much safer while performing transactions from their mobile devices.

### 5.1 Improvement Suggestions

1. Banking applications should use APK Signature Scheme V2 or V3 as it gives more protection against unauthorized alternations in APKs.
2. Banking applications should not be signed with SHA1with RSA. Since SHA1 is known to have collision issues. They should use SHA-256 or SHA-512 for better security.
3. Avoid printing logs, web-views, insecure functions, and encryption modes that lead to data leakage in the final release build of the banking application.
4. Use tools such as "SAST by KIUWAN" which helps developers to secure code during application development. SAST identifies and remediate vulnerabilities in code and makes it secure at every stage.
5. Do not store unnecessary information on the user device in any form.
6. Train and implement payment blocking model with the help of AI and ML that blocks payments that seem suspicious to the system.

### 5.2 Limitation of the Study

This study was carried out as part of an ongoing process of evaluating the security posture of mobile application throughout the state of Qatar, and to provide a benchmark framework for the evaluation of same. In this regard, there are some observable limitation in this current research output which will serve as input for further studies. These include:

- The unavailability of the credentials of M-Banking App-2 and M-Banking App-1 which prevents us from carrying out a dynamic analysis of both applications. A dynamic analysis has the capability to reveal further insight into the security posture of the evaluated application. Furthermore, such an analysis would provide a measure for simulating potential attack against observed vulnerabilities.
- Also, access to source code is not allowed which makes it difficult to do the low-level analysis of security at the code level.
- Unable to handshake backend database services which prevents the evaluation of authorization and authentication of both mobile applications.
- Another limitation of this study is the use of limited number of security evaluation tools, on limited number of mobile banking application. An exhaustive study would be required to draw a conclusion of the security posture of most mobile banking applications in the State. However, this study does provide a baseline on which further research can be developed. The exploration of forensics tools was not considered in this study. The integration of forensic tools in tandem with security tools have been asserted to provide a reliable metrics for evaluation a security status.

### 5.3 Potential Future Works

In future work, further analysis of these applications on other tools could open new dimensions for research in terms of security and helps to make these applications more secure. Such advanced studies can be used to extract the credentials of the applications (M-Banking App-1 and M-Banking App-2 in this case) which can help in doing dynamic analysis. Approaches to ensure collaboration with the banking organizations that can help to do a further security analysis of their applications to enhanced end-user safety. This approach can also help researchers to prevent false positives through heuristics from banking agencies. The development of a Pre- and Post-incident analysis framework is a potential direction for future studies. Particularly, leveraging studies that have explored forensic modalities [13]–[15] as well as machine learning alternatives [16][3], [17] towards enhancing the security analysis of mobile banking applications. In relation to global best practices for mobile application, a potential direction for future work is the development of a model for evaluating the security posture of a mobile application for the State of Qatar. Such a model would however consider the context-specific requirement of the State.

### 6. CONCLUSION

Exploring the security issues related to android mobile banking applications in the State of Qatar using static analysis approach was carried out in this study. A summary of the security scores for two mobile banking applications revealed a poor security practice which require urgent hotfixes. Some major security issues highlighted about both apps include the use of external storage for user sensitive data, a broadcast is sent without receiver permission, as well as vulnerability ECB cipher usage. These findings call for a more in-depth analysis of the entire m-baking ecosystem of the State of Qatar. Furthermore, a comprehensive model for ensuring user confidence in m-banking applications is essentially critical.